\documentclass[sigplan,screen,sigconf]{acmart}
\usepackage[]{hyperref}

\setcopyright{rightsretained}
\acmDOI{}
\acmISBN{}
\acmConference[LATTE '21]{1st Workshop on Languages, Tools, and Techniques for Accelerator Design}{April 15, 2021}{Virtual, Earth}

\begin{document}

\title{Elastic Silicon Interconnects}
\subtitle{Abstracting Communication in Accelerator Design}

\author{John Demme}
\email{john.demme@microsoft.com}
\affiliation{
  \institution{Microsoft, USA}
  \country{}
}

\begin{abstract}
  Communication is an important part of accelerator design, though it is
  under researched and under developed. Today, designers often face
  relatively low-level communication tools requiring them to design
  straightforward but error-prone plumbing. In this paper, we argue that
  raising the level of abstraction could yield correctness, productivity, and
  performance benefits not only for RTL-level designers but also for high
  level language developers.
\end{abstract}

\maketitle

\section{Introduction}

While there has been a huge amount of work in languages for hardware
accelerator design, the communication networks to which they are connected
has not seen much innovation lately. The euphemism ``plumbing'' is often used,
implying that the relatively mechanical task of moving data around to be a
painful, necessary evil. Today, it is. RTL designers not only have to
correctly implement a wire-level protocol, but correctly interpret the data
blob which sits on the wires. High level languages generally expose some
interfaces, but the data types on those ports are not known or enforced by
tooling. High level languages do help to make typed communication with the
host easy -- as long as the designer works in the same vendors environment
throughout the design on a board they support. In both cases, designers who
do this ``IP stitching'' must often ``gasket'' interfaces and convert untyped
data.

A high-level interconnect system should make these problems easier to solve.
It should provide a common, high-level type system to enable inter-language
communication and static type checking. It should automatically convert
to/from various signaling standards (instead of defining a new one) and
abstract away the details of wire-level signaling when the designer simply
does not care. It should make host-accelerator communicate trivial {\it
regardless} of the language, environment, or board the designer wants to use.
It should enable {\it incremental} adoption of new technologies, easing
integration of IP from an innovative tool. We could go on.

These problems motivated the Elastic Silicon Interconnects project
(henceforth called ESI). At its core, ESI is an elastic interconnect compiler
and a high-level type system. We then build on this to provide a rich set of
features.

\section{Types and elasticity}

ESI uses connectivity specifications which are
elastic\cite{carmona2009elastic} (based on
latency-insensitivity\cite{carloni2001theory}). This property provides the
flexibility ESI needs to handle relatively mundane tasks like automatically
pipelining connections, gearboxing (reducing or increasing the wire width of
a connection), translating signaling standards, and crossing clock domains.
Elasticity also allows ESI to route over other data transports, like PCIe to
a host CPU or over an IP network to either a host or another accelerator.

Streaming channels in ESI transmit and receive typed messages. These types
include the basic fixed size constructs: arbitrary width integers, enums,
arrays, unions, and structs. We also support variable-length types -- lists,
namely. Lists get transmitted in chunks over multiple cycles. Both the
transmitter and receiver will be allowed to select their own chunk sizes,
which may be different -- ESI will build the necessary gasket. Memory mapped
regions will be modeled by structs which can be written to and read from.

\begin{figure}[H]
\centering
\includegraphics[scale=0.65]{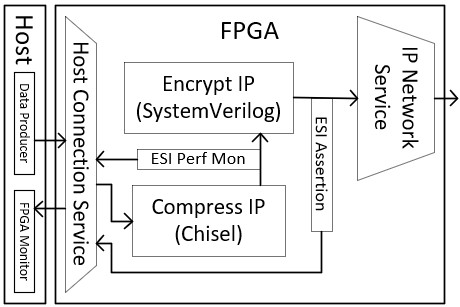}
\caption{An example ESI system with a performance monitor on the compressed
data stream and an assertion on the encrypted data stream.}
\end{figure}

\section{Composability}

ESI's elasticity and type information make systems easier to compose
correctly. For software integration, we can construct {\it typed} software
APIs for each connection. IP modules can be written in any language and be
attached to other parts of the system (including the host) as long as the
compiler speaks ESI or the designer specifies latency-insensitive
connections and types. In both cases, static type safety can be enforced by
the compiler.

These connections need not be to/from hardware either -- ESI has co-simulation
capabilities. One can use a generated API to talk to simulation then use the
same API when graduating to hardware. Alternatively, one could model an
ESI systems IP modules with software emulators then replace them gradually as
the hardware is developed. For longer IP module test runs, one could write
the test bench in software then synthesize the module to run faster on an
FPGA with the same API as in simulation.

\section{Futures: monitoring hooks}

Accelerator design and use requires monitoring in various forms. During
functional design, debugging is necessary and not fun. Staring at waveforms
from a simulation and/or guessing where to insert probes in synthesized
hardware is not productive. ESI can insert monitors on specified connections
and capture/transmit data only on valid messages vs every cycle. Since ESI
knows the types, graphical debuggers could use that information to display
debug information as intelligently and concisely as possible, either out of
simulation or in hardware (with some trade-offs). One could also imagine
inserting synthesizable assertions and being alerted on failure via a
low-overhead aggregation network which ESI could automatically plumb out.
When performance debugging, one could insert performance counters to get
aggregate data on messages passing through ESI connections.

In production use, telemetry becomes necessary yet often added as an
afterthought. It is often painful to plumb critical, low-bandwidth
information out of the design. ESI can help in two ways: First, it could
insert monitors of various types on its connections to monitor the
interconnect activity in production. Second, it could construct low-bandwidth
networks by serializing telemetry data into an arbitrarily low number of
wires, saving resources. The ease and relatively low overheads involved in
ESI telemetry should encourage more runtime telemetry earlier on to catch
issues which occur during use at scale and with
real workloads.

\section{Futures: services and board support}

The future feature which enables bridging arbitrary ESI modules to host software
and the monitoring hooks discussed above are ESI services. Any component
instantiated by ESI can request a connection to a service, which will in turn
provide some sort of specialized plumbing to said component. For instance, a
`telemetry' service could be provided to automatically wire up all the
components which provide telemetry data. Users don't usually care about the
speed or way that telemetry is hooked up, so long as it works and is low
overhead. Another example is an `assertion' service, which components could
use to alert a monitor that something has gone wrong, possibly with some
location and debug data. RTL modules and high level language modules could
also use services for common tasks, like low-bandwidth control plane
activities.

We can extend the notion of services to abstract away the board support
packages (a.k.a `shells') into a set of {\it standardized} board support
services. A `host communication' service would provide components the ability
to automatically connect to software, independent of the transport mechanism:
PCIe, ethernet, or cosim. That service would be implemented hierarchically
via a `PCIe' service, a `network' service, etc. Board support services could
even include `external memory' services to connect to off-chip DRAM.

\section{Related work}

\paragraph{Networks on Chip} When designers think of interconnects, they tend
to think of NoCs. Networks have become the standard for ASIC SoCs, and gain
popularity in FPGAs\cite{hilton2006pnoc, papamichael2012connect}, though
usually as a hardened component\cite{abdelfattah2013case} or in the control
plane. We, however, think of NoCs as a possible communication implementation.
ESI could synthesize a NoC or something ad-hoc depending on the connections
specified. If it or the designer chose a NoC, it could customize the topology
and/or only provide partial connectivity. We believe that starting from a
high level description of a designers requirements first then selecting an
implementation has benefits.

\paragraph{System Designers} Tools like Qsys and Vivado IP Integrator make
connectivity easier; however, they encourage vendor lock-in and remain low
level. Being reductive for space reasons, they are essentially graphical
NoC/SoC designers.

\paragraph{Integrated Development Environments} Commercial HLS environments
often provide an automated build flow to synthesize a full
software-to-hardware experience. They excel at integrating accelerators of a
particular type and specific language. These tools, however, make anything
that doesn't fit in that mold difficult: integrating IP from multiple
compilers, creating software APIs for unsupported
languages, porting to different systems (et cetera) all become more
difficult. ESI is just another tool -- it produces code to a system spec and
assumes that code complies with the spec.

\paragraph{Services and host communication} Many of the concepts we discuss
above were proposed and implemented in LEAP\cite{fleming2016leap}. They
implemented latency-insensitive channels which connect modules and use them
to partition designs across FPGAs and provide services. ESI builds on LEAP
(in an intellectual sense) in multiple dimensions. Sadly, LEAP never caught
on, is relatively inaccessible as it is written in Bluespec, and has not seen
any recent updates. Connectal\cite{king2015software} has an inteface
description language and it creates typed software-hardware RPC interfaces,
similar to the off-chip features of ESI.

\section{Going forward}

All of the benefits of ESI apply equally to both hand-coded RTL and high level
languages. A large part of our motivation is to enable both easier
development of languages and adoption. In the ESI world, high level compilers
are only responsible for spitting out ESI modules. Additionally, existing RTL
designs need not trust the new compilers with their whole design as they can
substitute their existing IP with the new one and compare. The intent is to
enable easier on-boarding of higher level languages by ensuring they play
nice with each other, the existing RTL, and the system.

ESI is a nascent project intended for production use and is part of the
\href{http://circt.org}{CIRCT project}. Today, it is capable of connecting
ESI ports on modules (with configurable buffering) to each other or to
cosimulation endpoints. Cosimulation communication is currently implemented
via Cap'nProto RPC\cite{capnp} so software drivers can be written in a
multitude of languages. The current implementation does not yet support
lists or gearboxing. Likewise, services need more development (both
intellectual and engineering). We have lofty goals which won't be achievable
without community involvement.

\bibliographystyle{ACM-Reference-Format}
\bibliography{master}

\end{document}